\documentclass[conference,final,10pt,a4paper]{IEEEtran}

\usepackage{cite}
\usepackage[pdftex]{graphicx}
 \graphicspath{{../pdf/}{../jpeg/}}
 \DeclareGraphicsExtensions{.pdf,.jpeg,.png}
\usepackage{pgf}
\usepackage{pgfplots}
\usepackage[cmex10]{amsmath}
\usepackage{amssymb}
\usepackage{algorithmic}
\usepackage{array}
\usepackage{mdwmath}
\usepackage{mdwtab}
\usepackage{url}
\usepackage{fixedmultirow}
\usepackage{trfsigns}
\usepackage{caption}
\usepackage{subcaption}
\usepackage{stfloats}
\usepackage{nicefrac}

\graphicspath{{graphics/}}
\graphicspath{{../Bilder/}}

\usetikzlibrary{calc,trees,positioning,arrows,chains,shapes.geometric,%
 decorations.pathreplacing,decorations.pathmorphing,shapes,%
 matrix,shapes.symbols,plotmarks,decorations.markings,shadows,fit,patterns}
\usetikzlibrary{spy,backgrounds}
\usepgfplotslibrary{groupplots}

\pgfplotsset{compat=1.3}
\tikzset{every picture/.style={>=latex}} 
\tikzset{every tick label/.style={font=\footnotesize}} 
\tikzset{every axis label/.style={font=\small}} 
\pgfsetplotmarksize{2pt}

\tikzset{orientation/.is choice,
    orientation/lr/.style={anchor=west,right=1},
    orientation/lr2/.style={anchor=west,right=2},
    orientation/lrd/.style={anchor=west,below=1},
    orientation/lrd2/.style={anchor=west,below=2},
    orientation/rl/.style={anchor=east,left=1},
    orientation/rl2/.style={anchor=east,left=2},
    orientation/ud/.style={anchor=north,below=1},
    orientation/du/.style={anchor=south,above=1},
    orientation/rld/.style={anchor=east,below=1},
    orientation/rld2/.style={anchor=east,below=2},
}
\tikzstyle{scare} = [
]
\tikzstyle{syslinear} = [
 drop shadow={shadow xshift=.6mm,
              shadow yshift=-.6mm},
 fill=white,
 anchor=west,
 rectangle,
 draw=black,
 minimum height=5mm,
 minimum width=5mm,
 inner xsep=0.5em
]

\tikzstyle{sysnonlinear} = [
 drop shadow={shadow xshift=.8mm,
              shadow yshift=-.8mm},
 fill=white,
 double,
 anchor=west,
 rectangle,
 draw=black,
 minimum height=5mm,
 minimum width=5mm,
 inner xsep=0.5em
]

\tikzstyle{syssource} = [
 anchor=west,
 ellipse,
 draw=black,
 minimum height=1.5ex,
 minimum width=1.5em,
 inner xsep=0.5em
]

\tikzstyle{syssink} = [
 anchor=west,
 ellipse,
 draw=black,
 minimum height=1.5ex,
 minimum width=1.5em,
 inner xsep=0.5em
]

\tikzstyle{syssplit} = [
 fill=black,
 draw=black,
]

\tikzstyle{sysadd} = [
 draw,circle,inner sep=-1pt,
]

\tikzstyle{sysmul} = [
 draw,circle,inner sep=-1pt,
]

\definecolor{MyHSBGreen}{hsb}{0.34065,1,0.91}
\pgfplotscreateplotcyclelist{colors4}{%
 {black!100!white},
 {black!75!white},
 {black!50!white},
 {black!25!white},
}
\pgfplotscreateplotcyclelist{colors5}{%
 {black!100!white},
 {black!75!white},
 {black!50!white},
 {black!25!white},
 {densely dashed, black!100!white},
 {densely dashed, black!75!white},
}
\pgfplotscreateplotcyclelist{colors10}{%
 {black!100!white},
 {black!75!white},
 {black!50!white},
 {black!25!white},
 {densely dashed, black!100!white},
 {densely dashed, black!75!white},
 {densely dashed, black!50!white},
 {densely dashed, black!25!white},
 {densely dotted, black!100!white},
 {densely dotted, black!75!white},
 {densely dotted, black!50!white},
}

\pgfplotscreateplotcyclelist{colorsBERMDDFSE}{%
                 every mark/.append style={fill=black},mark=o\\%
                 every mark/.append style={fill=black},mark=star\\%
                 every mark/.append style={fill=black},mark=triangle\\%
                 every mark/.append style={scale=.7,fill=black},mark=square\\%
                 every mark/.append style={fill=black},mark=diamond\\%
  densely dashed,every mark/.append style={solid},mark=pentagon*\\%
  densely dashed,every mark/.append style={solid},mark=*\\%
  densely dashed,every mark/.append style={solid},mark=triangle*\\%
  densely dashed,every mark/.append style={scale=.7,solid},mark=square*\\%
  densely dashed,every mark/.append style={solid},mark=diamond*\\%
  densely dashed,every mark/.append style={solid},mark=star*\\%
}
\pgfplotscreateplotcyclelist{colors1Empty4Full}{%
                 every mark/.append style={fill=black},mark=o\\%
  densely dashed,every mark/.append style={solid},mark=*\\%
  densely dashed,every mark/.append style={solid},mark=triangle*\\%
  densely dashed,every mark/.append style={scale=.7,solid},mark=square*\\%
  densely dashed,every mark/.append style={solid},mark=diamond*\\%
}

\pgfplotscreateplotcyclelist{colors4Empty4Full}{%
                 every mark/.append style={fill=black},mark=o\\%
                 every mark/.append style={fill=black},mark=triangle\\%
                 every mark/.append style={scale=.7,fill=black},mark=square\\%
                 every mark/.append style={fill=black},mark=diamond\\%
  densely dashed,every mark/.append style={solid},mark=*\\%
  densely dashed,every mark/.append style={solid},mark=triangle*\\%
  densely dashed,every mark/.append style={scale=.7,solid},mark=square*\\%
  densely dashed,every mark/.append style={solid},mark=diamond*\\%
}
\pgfplotscreateplotcyclelist{colors6Empty4Full}{%
                 every mark/.append style={fill=black},mark=o\\%
                 every mark/.append style={fill=black},mark=star\\%
                 every mark/.append style={fill=black},mark=triangle\\%
                 every mark/.append style={scale=.7,fill=black},mark=square\\%
                 every mark/.append style={fill=black},mark=diamond\\%
                 every mark/.append style={fill=black},mark=pentagon\\%
  densely dashed,every mark/.append style={solid},mark=*\\%
  densely dashed,every mark/.append style={solid},mark=triangle*\\%
  densely dashed,every mark/.append style={scale=.7,solid},mark=square*\\%
  densely dashed,every mark/.append style={solid},mark=diamond*\\%
}
\pgfplotscreateplotcyclelist{colors8Empty4Full}{%
                 every mark/.append style={fill=black},mark=o\\%
                 every mark/.append style={fill=black},mark=star\\%
                 every mark/.append style={fill=black},mark=triangle\\%
                 every mark/.append style={scale=.7,fill=black},mark=square\\%
                 every mark/.append style={fill=black},mark=diamond\\%
                 every mark/.append style={fill=black},mark=pentagon\\%
                 every mark/.append style={fill=black},mark=oplus\\%
                 every mark/.append style={fill=black},mark=otimes\\%
  densely dashed,every mark/.append style={solid},mark=*\\%
  densely dashed,every mark/.append style={solid},mark=triangle*\\%
  densely dashed,every mark/.append style={scale=.7,solid},mark=square*\\%
  densely dashed,every mark/.append style={solid},mark=diamond*\\%
}

\pgfplotscreateplotcyclelist{colors}{%
                 every mark/.append style={fill=black},mark=o\\%
                 every mark/.append style={fill=black},mark=star\\%
                 every mark/.append style={fill=black},mark=triangle\\%
                 every mark/.append style={scale=.7,fill=black},mark=square\\%
                 every mark/.append style={fill=black},mark=diamond\\%
                 every mark/.append style={fill=black},mark=pentagon\\%
                 every mark/.append style={fill=black},mark=oplus\\%
                 every mark/.append style={fill=black},mark=otimes\\%
                 every mark/.append style={fill=black},mark=asterisk\\%
                 every mark/.append style={fill=black},mark=+\\%
                 every mark/.append style={fill=black},mark=-\\%
                 every mark/.append style={fill=black},mark=|\\%
                 every mark/.append style={fill=black},mark=x\\%
                 every mark/.append style={fill=black},mark=*\\%
                 every mark/.append style={fill=black},mark=triangle*\\%
                 every mark/.append style={scale=.7,fill=black},mark=square*\\%
                 every mark/.append style={fill=black},mark=diamond*\\%
                 every mark/.append style={fill=black},mark=star*\\%
  densely dashed,every mark/.append style={solid,fill=gray},mark=o\\%
  densely dashed,every mark/.append style={solid,fill=gray},mark=triangle\\%
  densely dashed,every mark/.append style={scale=.7,solid,fill=gray},mark=square\\%
  densely dashed,every mark/.append style={solid,fill=gray},mark=diamond\\%
  densely dashed,every mark/.append style={solid,fill=gray},mark=star\\%
  densely dashed,every mark/.append style={solid,fill=gray},mark=pentagon*\\%
  densely dashed,every mark/.append style={solid,fill=gray},mark=*\\%
  densely dashed,every mark/.append style={solid,fill=gray},mark=triangle*\\%
  densely dashed,every mark/.append style={scale=.7,solid,fill=gray},mark=square*\\%
  densely dashed,every mark/.append style={solid,fill=gray},mark=diamond*\\%
  densely dashed,every mark/.append style={solid,fill=gray},mark=star*\\%
  mark=text,text mark=a\\%
  mark=text,text mark=b\\%
  mark=text,text mark=c\\%
  mark=text,text mark=d\\%
  mark=text,text mark=e\\%
  mark=text,text mark=f\\%
  mark=text,text mark=g\\%
}

\pgfdeclarelayer{background layer}
\pgfdeclarelayer{foreground layer}
\pgfsetlayers{background layer,main,foreground layer}

\newcommand{\ie}{\emph{i.e.}}
\newcommand{\eg}{\emph{e.g.}}
\newcommand{\cf}{\emph{cf.}}

\providecommand{\de}[1]{\ensuremath{\mathop{\mathrm{d}}}}

\tikzstyle{EMTY} =   [ fill=white, ]
\tikzstyle{STATE} =  [ fill=black!40!white, ]
\tikzstyle{INPUT} =  [ fill=black!20!white, ]
\tikzstyle{EXTEND} = [ pattern=north east lines, ]
\tikzstyle{XSB} =    [ thick, |-|, shorten <=3pt, shorten >=3pt ]

\IEEEoverridecommandlockouts
\title{Matched Decoding for Punctured Convolutional Encoded Transmission Over ISI-Channels}
\author{
 \IEEEauthorblockN{Fabian Schuh,
                   Andreas Schenk, and
                   Johannes B. Huber}%
 \IEEEauthorblockA{Institute for Information Transmission,
                   Friedrich-Alexander-Universit\"at Erlangen-N\"urnberg, Germany\\ 
                   mail: \texttt{\{schuh,\,schenk,\,huber\}@LNT.de}}%
 \thanks{This work was supported by Bundesministerium f\"ur Wirtschaft und
         Technologie (BMWi) within the project C-PMSE.}
}

\begin{document}
\maketitle
\begin{abstract}
 Matched decoding is a technique that enables the efficient maximum-likelihood
 sequence estimation of convolutionally encoded PAM-transmission over
 ISI-channels. Recently, we have shown that the super-trellis of encoder and
 channel can be described with significantly fewer states without loss in
 Euclidean distance, by introducing a non-linear representation of the trellis.
 This paper extends the matched decoding concept to punctured convolutional
 codes and introduces a time-variant, non-linear trellis description.
\end{abstract}
\begin{IEEEkeywords}
 ISI-channel;
 punctured convolutionally encoded transmission;
 super-trellis decoding;
 matched decoding;
\end{IEEEkeywords}
\IEEEpeerreviewmaketitle
\section{Introduction}\label{sec:intro}

Coded pulse-amplitude modulation poses an attractive digital transmission
scheme when low over-all delay is desired but the channel induces intersymbol
interference (ISI). Low latency is obtained by the use of convolutional codes
instead of block codes (\cf~\cite{LIT_tr_com_2009_hehn}) and dispense with
interleaving (as opposed to convolutional bit-interleaved coded
modulation~\cite{141453}).

Punctured codes are widely used in applications where high code rates are
required. With puncturing, a convolutional code with high code rate can be
derived from \emph{mother code} with a low code rate.
At the receiver, equalization and decoding are usually performed subsequently
in two separate processing steps, each based on its own trellis description.
After equalization,\footnote{The bit metrics of punctured symbols are set to an
erasure (\eg, $\mathcal{L}=0$) before decoding.} bit probability $\nicefrac12$
are inserted before decoding when symbols are punctured.


In~\cite{2012arXiv1207.4680S} we proposed a matched decoding scheme where
super-trellis decoding of the ISI-channel and \emph{non-punctured}
convolutional encoding is performed jointly on a reduced number of states.
There, the output of the rate-$\nicefrac{K}{n}$ convolutional encoder was mapped
onto $M=2^n$ symbols directly following the concept of trellis coded
modulation.
In our setup, we will perform equalization of the ISI-channel and decoding of
the punctured convolutional code jointly in a single
super-trellis~\cite{huber1992trelliscodierung}. This technique, however, is
commonly not applied practically due to the large overall number of states of a
super-trellis as well as the complex finite state machine to represent the
encoding with a punctured convolutional code.
However, we show how to extend the matched decoding (MD) concept to the use
with punctured convolutional codes and therefore reduce the total number of
states. For this, we will first show how to describe the encoding with a
punctured code as a finite state machine (FSM), only, and extend it to
compute the hypotheses for the punctured convolutional code transmitted
over an ISI-channel. We will show that the trellis described by the extended
FSM is time-variant and can be used for decoding using a modified version of the
Viterbi algorithm~(VA)~\cite{1054010}.

The paper is organized as follows: After the definition of the system model for
punctured convolutional coded transmission over ISI-channels in
Sec.~\ref{sec:systemModel}, we will first describe the FSM for
\emph{non-punctured} convolutional codes and briefly recall matched decoding in
Sec.~\ref{sec:MDforNonPunct}. In Sec.~\ref{sec:punctMatched} we will describe
the FSM for punctured codes and extend our MD approach. Numerical simulations
are shown in Sec.~\ref{sec:numericalResults}. The paper concludes with a
summary.

\section{System Model}\label{sec:systemModel}

\begin{figure}[ht]\vspace*{-2ex}
 \begin{center}
  \begin{tikzpicture}[>=latex,x=1em,y=4ex,font=\footnotesize,inner sep=0.3em,
                      node distance=10mm and 4mm]
   \node at (0,0) (u) {$u[k]$};
   \node[coordinate,right=of u] (in) {};
   \draw node[syslinear,right=of in] (T1) {};
   \draw node[syslinear,right=of T1] (T2) {};
   \draw node[sysadd, above=of T1.east,xshift=2mm,rectangle,inner sep=0pt] (g1) {$+$};
   \draw node[sysadd, below=of T1.east,xshift=2mm,rectangle,inner sep=0pt] (g2) {$+$};
   \node[anchor=south,at=(g1.north)] {$\operatorname{mod}2$};
   \node[anchor=north,at=(g2.south)] {$\operatorname{mod}2$};
   \path (u)       edge[o->] node[coordinate,pos=0.6] (u1) {} (T1)
         (T1)      edge[->]  node[coordinate,midway] (u2) {} (T2)
         (T2.east) edge[-]   node[coordinate,pos=1] (u3) {}  ++(2.5mm,0);
   \draw[fill]  (u1) circle(.5pt) -- ++(0,5mm) edge[->] (g1)
                (u3) circle(.5pt) -- ++(0,5mm) edge[->] (g1)
                (u1) circle(.5pt) -- ++(0,-5mm) edge[->] (g2)
                (u2) circle(.5pt) -- ++(0,-5mm) edge[->] (g2)
                (u3) circle(.5pt) -- ++(0,-5mm) edge[->] (g2);
   \node[syslinear,right=10mm,at=(T2.east)] (Mapper) {Mapper};
   \draw[->] (g1) to[->] ++(7mm,0) node[sysnonlinear,inner xsep=2pt] (p1) {$P_1\!:\!\left\{ 1\;\;0 \right\}$};
   \draw[->] (g2) to[->] ++(7mm,0) node[sysnonlinear,inner xsep=2pt] (p2) {$P_2\!:\!\left\{ 1\;\;1 \right\}$};
   \draw[->] (p1.south) |- ($(Mapper.west)+(0,3pt)$);
   \draw[->] (p2.north) |- ($(Mapper.west)-(0,3pt)$);
   \draw node[syslinear,at=(Mapper),xshift=15mm] (T3) {};
   \draw node[syslinear,right=of T3] (T4) {};
   \draw (Mapper)   edge[->] node[coordinate,midway] (u4) {} (T3)
         (T3)       edge[->] node[coordinate,midway] (u5) {} (T4)
         (T4.east)  edge[-] node[coordinate,pos=1]   (u6) {} ++(2.5mm,0);
   \node[above] at (u4) {$m[k]$};
   \node[sysmul,at=(u4),yshift=-10mm] (h1) {$\times$};
   \node[sysmul,at=(u5),yshift=-10mm] (h2) {$\times$};
   \node[sysmul,at=(u6),yshift=-10mm] (h3) {$\times$};
   \draw[<-] (h1) -- ++(-5mm,0) node[pos=0.5,anchor=south] {$h[0]$};
   \draw[<-] (h2) -- ++(-5mm,0) node[pos=0.5,anchor=south] {$h[1]$};
   \draw[<-] (h3) -- ++(-5mm,0) node[pos=0.5,anchor=south] {$h[2]$};
    \draw node[sysadd, below=of h2.south,yshift=5mm] (h) {$+$};
   \draw[fill] (u4) circle(.5pt) -- (h1) edge[->] (h)
               (u5) circle(.5pt) -- (h2) edge[->] (h)
               (u6) circle(.5pt) -- (h3) edge[->] (h);
   \draw[-o] (h) -- ++(5mm,0) node[right] {$r[k]$};
  \end{tikzpicture}\vspace*{-3ex}
 \end{center}
 \caption{System model for a encoded $4$-ary transmission
          over ISI-channels with a punctured convolutional code and overall
          rate $R=\nicefrac43$.}
 \label{fig:sysmodel}
 \vspace*{-2ex}
\end{figure}
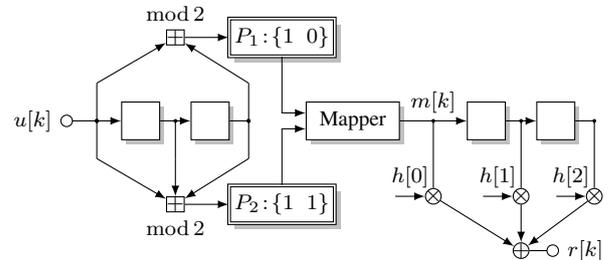

We first introduce the system model for a convolutionally encoded PAM
transmission over ISI-channel (\cf\ discrete-time example of
Fig.~\ref{fig:sysmodel}).

The transmitter is composed of a rate-$\nicefrac{K}{n}$ binary convolutional
encoder with generator polynomials $\left[\,g_{ij}\,\right]$; \mbox{$1\leq i\leq
n$}; \mbox{$1 \leq j\leq K$}, with $K$ input symbols and $n$ parallel output
symbols at each time instant. At each output branch of the encoder, the symbols
traverses through a non-linear time-invariant puncturing system with periodic
puncturing scheme $P_i;\;1<i\leq n$. For each encoder input symbol the
puncturing scheme cyclically advances by one step. Whenever the scheme is zero,
the current symbol at the output gets punctured (\ie, discarded), accordingly.
The punctured output is mapped for a $M$-ary PAM transmission. The transmit
signal traverses through a memory-$L$ discrete-time ISI-channel with $L+1$
channel coefficients $h[k]$ with $k$ denoting the time index.

The only requirement necessary for the matched decoding approach is that the
rate-$\nicefrac{K}{n}$ convolutional code is matched to the $M$-ary modulation via
$M=2^n$. For sake of simplicity, we here consider real-valued ASK, only. For
clarity, we restrict ourselves to $M=4$, but note that the concept can be
extended to arbitrary $M=2^n$.

\section{Matched Decoding of Non-Punctured Codes}\label{sec:MDforNonPunct}

In order to introduce maximum-likelihood sequence estimation (MLSE) for
punctured codes, we first describe the encoding and mapping process for
\emph{non-punctured} convolutional encoding and natural labeling. We first
ignore the ISI-channel for simplicity, \ie, $L=0$, and later briefly recall our
MD approach also considering an ISI-channel. Please note that here a single
uncoded bit is encoded with a rate-$\nicefrac12$ convolutional encoder
resulting in two coded bits, \ie, MSB and LSB, respectively. These coded bits
contain the information of a single uncoded bit and are mapped onto a single
transmit symbol $m[k]$. Thus, after mapping, the overall rate of the
transmission is $R=1\nicefrac{\text{bit}}{\text{symbol}}$. 

\subsection{Description of the Finite State Machine}\label{sec:NonpunctCodesFSM}

Fig.~\ref{fig:convEncodingTranditional} illustrates the encoding process (top)
and the state transitions of the FSM (bottom) when no puncturing is involved.
The uncoded unipolar information sequence $u[k]\in\left\{ 0,\;1 \right\}$ is
inserted into the FSM (light gray, \tikz \draw[INPUT,baseline] rectangle
(1.5ex,1.5ex);\,) as input values and later passes through all delay elements
describing the state of the FSM (dark gray, \tikz \draw[STATE,baseline]
rectangle (1.5ex,1.5ex);\,). Here, the generator polynomials $g_i$, $1<i<2$,
process the input symbol together with the FSM state synchronously at each time
instant. The resulting encoded bits, denoted with MSB and LSB, respectively,
are naturally labeld and mapped to the $4$-ary symbol alphabet of the
transmission scheme, \eg, via $m = 2(2\text{MSB}+\text{LSB})-1$.

\begin{figure}[ht]\vspace*{-3ex}
 \begin{center}
  \begin{tikzpicture}[>=latex,x=6em,y=4ex,font=\footnotesize,inner sep=0.3em,
                      node distance=10mm and 4mm]
   \coordinate (u0) at (0,0);
   \coordinate (u1) at (1,0);
   \coordinate (u2) at (2,0);
   \coordinate (u3) at (3,0);
   \draw (u0) circle (1pt) node[above] {$u[\nu]$};
   \draw (u1) circle (1pt) node[above] {$u[\nu+1]$};
   \draw (u2) circle (1pt) node[above] {$u[\nu+2]$};
   \draw (u3) circle (1pt) node[above] {$u[\nu+3]$};
   \draw (-0.25,-1) circle (1pt) node[coordinate] (c0) {};
   \draw (+0.25,-1) circle (1pt) node[coordinate] (c1) {};
   \draw (+0.75,-1) circle (1pt) node[coordinate] (c2) {};
   \draw (+1.25,-1) circle (1pt) node[coordinate] (c3) {};
   \draw (+1.75,-1) circle (1pt) node[coordinate] (c4) {};
   \draw (+2.25,-1) circle (1pt) node[coordinate] (c5) {};
   \draw (+2.75,-1) circle (1pt) node[coordinate] (c6) {};
   \draw (+3.25,-1) circle (1pt) node[coordinate] (c7) {};
   \begin{scope}[shorten <=1pt,shorten >=1pt]
    \draw[->] (u0) -- node[midway,above,sloped] {$g_1$} (c0);
    \draw[->] (u0) -- node[midway,above,sloped] {$g_2$} (c1);
    \draw[->] (u1) -- node[midway,above,sloped] {$g_1$} (c2);
    \draw[->] (u1) -- node[midway,above,sloped] {$g_2$} (c3);
    \draw[->] (u2) -- node[midway,above,sloped] {$g_1$} (c4);
    \draw[->] (u2) -- node[midway,above,sloped] {$g_2$} (c5);
    \draw[->] (u3) -- node[midway,above,sloped] {$g_1$} (c6);
    \draw[->] (u3) -- node[midway,above,sloped] {$g_2$} (c7);
    \draw (-0.25,-2) node[draw] (s0) {MSB};
    \draw (+0.25,-2) node[draw] (s1) {LSB};
    \draw (+0.75,-2) node[draw] (s2) {MSB};
    \draw (+1.25,-2) node[draw] (s3) {LSB};
    \draw (+1.75,-2) node[draw] (s4) {MSB};
    \draw (+2.25,-2) node[draw] (s5) {LSB};
    \draw (+2.75,-2) node[draw] (s6) {MSB};
    \draw (+3.25,-2) node[draw] (s7) {LSB};
   \end{scope}
   \begin{scope}[shorten <=1pt,shorten >=1pt]
    \foreach \x in {0,...,7} {
     \draw[->] (c\x) -- (s\x);
    }
   \end{scope}
   \draw ($(s0.south west)-(1pt,1pt)$) -| ($(s1.north east)+(1pt,1pt)$) -| ($(s0.south west)-(1pt,1pt)$);
   \draw ($(s2.south west)-(1pt,1pt)$) -| ($(s3.north east)+(1pt,1pt)$) -| ($(s2.south west)-(1pt,1pt)$);
   \draw ($(s4.south west)-(1pt,1pt)$) -| ($(s5.north east)+(1pt,1pt)$) -| ($(s4.south west)-(1pt,1pt)$);
   \draw ($(s6.south west)-(1pt,1pt)$) -| ($(s7.north east)+(1pt,1pt)$) -| ($(s6.south west)-(1pt,1pt)$);
   \begin{scope}[decoration={brace,amplitude=.5em},decorate]
    \draw[decorate] ($(s1.south east)-(0,2pt)$) -- node[yshift=-1ex,midway,below] {$m[\nu]$} ($(s0.south west)-(0,2pt)$);
    \draw[decorate] ($(s3.south east)-(0,2pt)$) -- node[yshift=-1ex,midway,below] {$m[\nu+1]$} ($(s2.south west)-(0,2pt)$);
    \draw[decorate] ($(s5.south east)-(0,2pt)$) -- node[yshift=-1ex,midway,below] {$m[\nu+2]$} ($(s4.south west)-(0,2pt)$);
    \draw[decorate] ($(s7.south east)-(0,2pt)$) -- node[yshift=-1ex,midway,below] {$m[\nu+3]$} ($(s6.south west)-(0,2pt)$);
   \end{scope}
  \end{tikzpicture}
  \vspace{2ex}
  \begin{tikzpicture}
   \tikzset{row 1 column 5/.style={nodes={STATE}}}
   \tikzset{row 1 column 4/.style={nodes={STATE}}}
   \tikzset{row 1 column 3/.style={nodes={INPUT}}}
   \tikzset{row 2 column 4/.style={nodes={STATE}}}
   \tikzset{row 2 column 3/.style={nodes={STATE}}}
   \tikzset{row 2 column 2/.style={nodes={INPUT}}}
   \tikzset{row 3 column 3/.style={nodes={STATE}}}
   \tikzset{row 3 column 2/.style={nodes={STATE}}}
   \tikzset{row 3 column 1/.style={nodes={INPUT}}}
   \matrix (FIFO) [matrix of nodes,
                   nodes in empty cells,
                   nodes={draw,
                          ultra thin,
                          anchor=south,
                          rectangle,
                          text width=2em,
                          minimum height=7ex,
                         },
                   ] {
     &&&&\\
     &&&&\\
     &&&&\\
    };
    \draw[XSB] ($(FIFO-1-3.west)+(0,1ex)$) -- node[above,font=\footnotesize,midway] {$g_1$} ($(FIFO-1-5.east)+(0,1ex)$);
    \draw[XSB] ($(FIFO-1-3.west)-(0,1ex)$) -- node[below,font=\footnotesize,midway] {$g_2$} ($(FIFO-1-5.east)-(0,1ex)$);
    \draw[XSB] ($(FIFO-2-2.west)+(0,1ex)$) -- node[above,font=\footnotesize,midway] {$g_1$} ($(FIFO-2-4.east)+(0,1ex)$);
    \draw[XSB] ($(FIFO-2-2.west)-(0,1ex)$) -- node[below,font=\footnotesize,midway] {$g_2$} ($(FIFO-2-4.east)-(0,1ex)$);
    \draw[XSB] ($(FIFO-3-1.west)+(0,1ex)$) -- node[above,font=\footnotesize,midway] {$g_1$} ($(FIFO-3-3.east)+(0,1ex)$);
    \draw[XSB] ($(FIFO-3-1.west)-(0,1ex)$) -- node[below,font=\footnotesize,midway] {$g_2$} ($(FIFO-3-3.east)-(0,1ex)$);
    \node[font=\footnotesize,right=1pt,draw,inner sep=1pt] at ($(FIFO-1-5.east)+(0,1.5ex)$)  {MSB};
    \node[font=\footnotesize,right=1pt,draw,inner sep=1pt] at ($(FIFO-1-5.east)+(0,-1.5ex)$) {LSB};
    \node[font=\footnotesize,right=1pt,draw,inner sep=1pt] at ($(FIFO-2-4.east)+(0,1.5ex)$)  {MSB};
    \node[font=\footnotesize,right=1pt,draw,inner sep=1pt] at ($(FIFO-2-4.east)+(0,-1.5ex)$) {LSB};
    \node[font=\footnotesize,right=1pt,draw,inner sep=1pt] at ($(FIFO-3-3.east)+(0,1.5ex)$)  {MSB};
    \node[font=\footnotesize,right=1pt,draw,inner sep=1pt] at ($(FIFO-3-3.east)+(0,-1.5ex)$) {LSB};
    \node[above,font=\footnotesize] at (FIFO-1-1.north) {$\nu+2$};
    \node[above,font=\footnotesize] at (FIFO-1-2.north) {$\nu+1$};
    \node[above,font=\footnotesize] at (FIFO-1-3.north) {$\nu$};
    \node[above,font=\footnotesize] at (FIFO-1-4.north) {$\nu-1$};
    \node[above,font=\footnotesize] at (FIFO-1-5.north) {$\nu-2$};
  \end{tikzpicture}\vspace*{-4ex}
 \end{center}
 \caption{Top: Encoding process for a rate-$\nicefrac12$ convolutional code and
          $4$-ary mapping. Overall transmission rate $R=1$. Bottom:
          State transitions of the transmitter FSM and the relations between
          generator polynomials and FSM-state/input.}
 \label{fig:convEncodingTranditional}
 \vspace*{-2ex}
\end{figure}

The resulting trellis is well-known and time-invariant as in each time step the
same relation between input value, FSM state, and generator polynomials holds.

\subsection{Matched Decoding}\label{sec:NonpunctCodesME}

\begin{figure}[ht]\vspace*{-2ex}
 \begin{center}
  \begin{tikzpicture}[>=latex,x=10em,y=4ex,font=\footnotesize,inner sep=0.3em,
                      node distance=10mm and 4mm]
    \node at (0,0) (u) {};
    \node[coordinate,right=of u] (in) {};
    \draw node[syslinear,right=of in,xshift=-3mm] (T1) {};
    \draw node[syslinear,right=of T1] (T2) {};
    \draw node[sysadd, above=of T1.east,xshift=2mm] (g1) {$+$};
    \draw node[sysadd, below=of T1.east,xshift=2mm] (g2) {$+$};
    \path (u)       edge[o->] node[coordinate,pos=0.6] (u1) {} (T1)
          (T1)      edge[->]  node[coordinate,midway] (u2) {} (T2)
          (T2.east) edge[-]   node[coordinate,pos=1] (u3) {}  ++(2.0mm,0);
    \draw[fill]  (u1) circle(.5pt) -- ++(0,5mm) edge[->] (g1)
                 (u3) circle(.5pt) -- ++(0,5mm) edge[->] (g1)
                 (u1) circle(.5pt) -- ++(0,-5mm) edge[->] (g2)
                 (u2) circle(.5pt) -- ++(0,-5mm) edge[->] (g2)
                 (u3) circle(.5pt) -- ++(0,-5mm) edge[->] (g2);
    \node[sysnonlinear,at=(g1),xshift=12mm] (floorMSB) {$\left\lfloor\frac{\cdot}{2}\right\rfloor$};
    \draw[->] (g1) -- node[coordinate,pos=0.8] (preFloorMSB) {} (floorMSB);
    \node[sysmul,right=of floorMSB,xshift=-3mm] (MSBfloor2) {$\times$};
    \node[sysadd,right=of MSBfloor2,xshift=-2mm] (MSBfloorAdd) {$+$};
    \draw[<-] (MSBfloor2) -- ++(0,5mm) node[anchor=south] {$2$};
    \draw[->] (floorMSB) -- (MSBfloor2) -- (MSBfloorAdd) node[below left] {$-$};
    \draw[->] (preFloorMSB) -- ++(0,10mm) -| (MSBfloorAdd);
    \draw[fill] (preFloorMSB) circle(.5pt);
    \draw node[syslinear,at=(MSBfloorAdd),xshift=5mm] (T3MSB) {};
    \draw node[syslinear,right=of T3MSB] (T4MSB) {};
    \path (MSBfloorAdd) edge[->] node[coordinate,midway] (u4) {} (T3MSB)
          (T3MSB)       edge[->] node[coordinate,midway] (u5) {} (T4MSB)
          (T4MSB.east)  edge[-] node[coordinate,pos=1]   (u6) {} ++(2.5mm,0);
    \node[sysmul,at=(u4),yshift=-8mm] (MSBh1) {$\times$};
    \node[sysmul,at=(u5),yshift=-8mm] (MSBh2) {$\times$};
    \node[sysmul,at=(u6),yshift=-8mm] (MSBh3) {$\times$};
    \draw[<-] (MSBh1) -- ++(-5mm,0) node[pos=0.5,anchor=south] {$h[0]$};
    \draw[<-] (MSBh2) -- ++(-5mm,0) node[pos=0.5,anchor=south] {$h[1]$};
    \draw[<-] (MSBh3) -- ++(-5mm,0) node[pos=0.5,anchor=south] {$h[2]$};
    \draw node[sysadd, below=of MSBh2.south,yshift=5mm] (MSBh) {$+$};
    \draw[fill] (u4) circle(.5pt) -- (MSBh1) edge[->] (MSBh)
                (u5) circle(.5pt) -- (MSBh2) edge[->] (MSBh)
                (u6) circle(.5pt) -- (MSBh3) edge[->] (MSBh);
    \node[sysmul,at=(MSBh),xshift=12mm] (MSBh2) {$\times$};
    \draw[<-] (MSBh) -- (MSBh2) -- ++(0,3mm) node[anchor=south] {$2$};
    \node[sysnonlinear,at=(g2),xshift=12mm] (floorLSB) {$\left\lfloor\frac{\cdot}{2}\right\rfloor$};
    \draw[->] (g2) -- node[coordinate,pos=0.8] (preFloorLSB) {} (floorLSB);
    \node[sysmul,right=of floorLSB,xshift=-3mm] (LSBfloor2) {$\times$};
    \node[sysadd,right=of LSBfloor2,xshift=-2mm] (LSBfloorAdd) {$+$};
    \draw[<-] (LSBfloor2) -- ++(0,5mm) node[anchor=south] {$2$};
    \draw[->] (floorLSB) -- (LSBfloor2) -- (LSBfloorAdd) node[below left] {$-$};
    \draw[->] (preFloorLSB) -- ++(0,10mm) -| (LSBfloorAdd);
    \draw[fill] (preFloorLSB) circle(.5pt);
    \draw node[syslinear,at=(LSBfloorAdd),xshift=5mm] (T3LSB) {};
    \draw node[syslinear,right=of T3LSB] (T4LSB) {};
    \path (LSBfloorAdd) edge[->] node[coordinate,midway] (u4) {} (T3LSB)
          (T3LSB)       edge[->] node[coordinate,midway] (u5) {} (T4LSB)
          (T4LSB.east)  edge[-] node[coordinate,pos=1]   (u6) {} ++(2.5mm,0);
    \node[sysmul,at=(u4),yshift=-8mm] (LSBh1) {$\times$};
    \node[sysmul,at=(u5),yshift=-8mm] (LSBh2) {$\times$};
    \node[sysmul,at=(u6),yshift=-8mm] (LSBh3) {$\times$};
    \draw[<-] (LSBh1) -- ++(-5mm,0) node[pos=0.5,anchor=south] {$h[0]$};
    \draw[<-] (LSBh2) -- ++(-5mm,0) node[pos=0.5,anchor=south] {$h[1]$};
    \draw[<-] (LSBh3) -- ++(-5mm,0) node[pos=0.5,anchor=south] {$h[2]$};
    \draw node[sysadd, below=of LSBh2.south,yshift=5mm] (LSBh) {$+$};
    \draw[fill] (u4) circle(.5pt) -- (LSBh1) edge[->] (LSBh)
                (u5) circle(.5pt) -- (LSBh2) edge[->] (LSBh)
                (u6) circle(.5pt) -- (LSBh3) edge[->] (LSBh);
    \node[sysadd,at=(T2),xshift=49mm,yshift=-10mm] (sumAll) {$+$};
    \draw[->] (LSBh) -| (sumAll);
    \draw[->] (MSBh2) -| (sumAll);
    \node[sysmul,right=of sumAll,xshift=-3mm] (alltimes) {$\times$};
    \node[sysadd,right=of alltimes,xshift=-3mm] (allplus)  {$+$};
    \draw[<-] (alltimes) -- ++(0,5mm) node[anchor=south] {$2$};
    \draw[<-] (allplus) -- ++(0,5mm) node[anchor=south] {$C$};
    \draw[-o] (sumAll) -- (alltimes) -- (allplus) -- ++(4mm,0) node[right] {$r[k]$};
  \end{tikzpicture}\vspace{-2ex}
 \end{center}
 \caption{Replacement of the $\operatorname{mod}2$ addition with the
          non-linear representation using $\operatorname{floor}$
          function. ($M=4$, \ie, $n=2$, \emph{non-punctured} code).}
 \label{fig:EncStep3NonCohCPM}
 \vspace*{-2ex}
\end{figure}
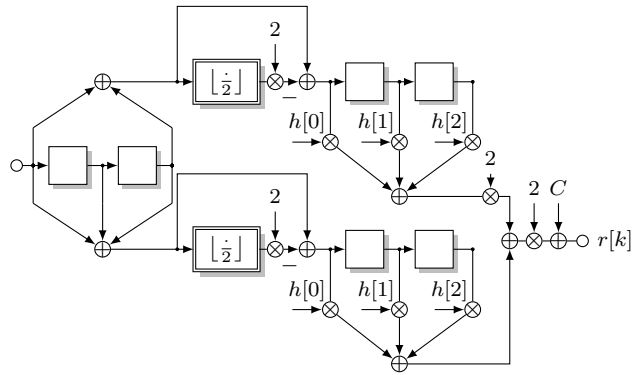

In the following, we recall the matched decoding approach briefly. As we
derived in~\cite{2012arXiv1207.4680S}, the matched decoding approach combines
the channel encoder with the ISI-channel in such a way, that fewer states are
sufficient to describe the super-trellis. There, the rate-$\nicefrac{K}{n}$
convolutional code was matched to a $2^n$-ary ASK-transmission scheme. Instead
of creating the super-trellis based on the $M$-ary channel, we represented the
super-trellis based on $\log_2{M}=n$ parallel binary ISI-channels. With $C =
-\sum\nolimits_{k=0}^{L}h[k](M-1)$ and the Gauss representation of the modulo
operation the transmission can be represented as depicted in
Fig.~\ref{fig:EncStep3NonCohCPM}. There, the memory of the ISI-channel is
binary only and a natural labeling is performed. This approach enables a first
state reduction of the super-trellis without loss in minimum Euclidean
distance.

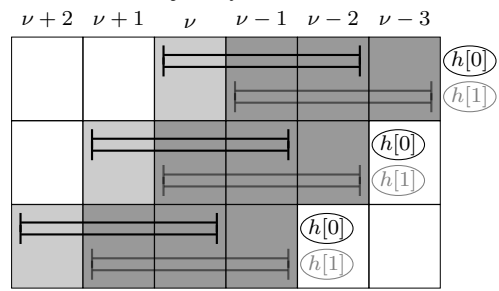
\begin{figure}[ht]\vspace*{-3ex}
 \begin{center}
  \begin{tikzpicture}
   \tikzset{row 1 column 3/.style={nodes={INPUT}}}
   \tikzset{row 1 column 4/.style={nodes={STATE}}}
   \tikzset{row 1 column 5/.style={nodes={STATE}}}
   \tikzset{row 1 column 6/.style={nodes={STATE}}}
   \tikzset{row 2 column 2/.style={nodes={INPUT}}}
   \tikzset{row 2 column 3/.style={nodes={STATE}}}
   \tikzset{row 2 column 4/.style={nodes={STATE}}}
   \tikzset{row 2 column 5/.style={nodes={STATE}}}
   \tikzset{row 3 column 1/.style={nodes={INPUT}}}
   \tikzset{row 3 column 2/.style={nodes={STATE}}}
   \tikzset{row 3 column 3/.style={nodes={STATE}}}
   \tikzset{row 3 column 4/.style={nodes={STATE}}}
   \matrix (FIFO) [matrix of nodes,
                   nodes in empty cells,
                   nodes={draw,
                          ultra thin,
                          anchor=south,
                          rectangle,
                          text width=2em,
                          minimum height=7ex,
                         },
                   ] {
     &&&&&\\
     &&&&&\\
     &&&&&\\
    };
    \draw[XSB] ($(FIFO-1-3.west)+(0,2ex)$) -- ($(FIFO-1-5.east)+(0,2ex)$);
    \draw[XSB] ($(FIFO-1-3.west)+(0,1ex)$) -- ($(FIFO-1-5.east)+(0,1ex)$);
    \draw[XSB,opacity=.5] ($(FIFO-1-4.west)-(0,1ex)$) -- ($(FIFO-1-6.east)-(0,1ex)$);
    \draw[XSB,opacity=.5] ($(FIFO-1-4.west)-(0,2ex)$) -- ($(FIFO-1-6.east)-(0,2ex)$);
    \draw[XSB] ($(FIFO-2-2.west)+(0,2ex)$) -- ($(FIFO-2-4.east)+(0,2ex)$);
    \draw[XSB] ($(FIFO-2-2.west)+(0,1ex)$) -- ($(FIFO-2-4.east)+(0,1ex)$);
    \draw[XSB,opacity=.5] ($(FIFO-2-3.west)-(0,1ex)$) -- ($(FIFO-2-5.east)-(0,1ex)$);
    \draw[XSB,opacity=.5] ($(FIFO-2-3.west)-(0,2ex)$) -- ($(FIFO-2-5.east)-(0,2ex)$);
    \draw[XSB] ($(FIFO-3-1.west)+(0,2ex)$) -- ($(FIFO-3-3.east)+(0,2ex)$);
    \draw[XSB] ($(FIFO-3-1.west)+(0,1ex)$) -- ($(FIFO-3-3.east)+(0,1ex)$);
    \draw[XSB,opacity=.5] ($(FIFO-3-2.west)-(0,1ex)$) -- ($(FIFO-3-4.east)-(0,1ex)$);
    \draw[XSB,opacity=.5] ($(FIFO-3-2.west)-(0,2ex)$) -- ($(FIFO-3-4.east)-(0,2ex)$);
    \node[font=\footnotesize,right=1pt,ellipse,draw,inner sep=0pt] at ($(FIFO-1-6.east)+(0,1.5ex)$) {$h[0]$};
    \node[font=\footnotesize,right=1pt,ellipse,draw,inner sep=0pt,opacity=.5] at ($(FIFO-1-6.east)+(0,-1.5ex)$) {$h[1]$};
    \node[font=\footnotesize,right=1pt,ellipse,draw,inner sep=0pt] at ($(FIFO-2-5.east)+(0,1.5ex)$) {$h[0]$};
    \node[font=\footnotesize,right=1pt,ellipse,draw,inner sep=0pt,opacity=.5] at ($(FIFO-2-5.east)+(0,-1.5ex)$) {$h[1]$};
    \node[font=\footnotesize,right=1pt,ellipse,draw,inner sep=0pt] at ($(FIFO-3-4.east)+(0,1.5ex)$) {$h[0]$};
    \node[font=\footnotesize,right=1pt,ellipse,draw,inner sep=0pt,opacity=.5] at ($(FIFO-3-4.east)+(0,-1.5ex)$) {$h[1]$};
    \node[above,font=\footnotesize] at (FIFO-1-1.north) {$\nu+2$};
    \node[above,font=\footnotesize] at (FIFO-1-2.north) {$\nu+1$};
    \node[above,font=\footnotesize] at (FIFO-1-3.north) {$\nu$};
    \node[above,font=\footnotesize] at (FIFO-1-4.north) {$\nu-1$};
    \node[above,font=\footnotesize] at (FIFO-1-5.north) {$\nu-2$};
    \node[above,font=\footnotesize] at (FIFO-1-6.north) {$\nu-3$};
  \end{tikzpicture}\vspace{-2ex}
 \end{center}
 \caption{State transitions of the FSM for matched decoding.}
 \label{fig:FSMMatchedNonPunct}
 \vspace*{-2ex}
\end{figure}

Figure~\ref{fig:FSMMatchedNonPunct} illustrates the state transitions of the
transmitter FSM when \emph{non-punctured} convolutional encoding and
ISI-channel are considered jointly.

\section{Matched Decoding for Punctured Codes}\label{sec:punctMatched}

We now describe the encoding process and state transitions when \emph{punctured
codes} are used. We will discuss the FSM for punctured convolutional codes as
well as the modifications necessary for the VA to run in the resulting trellis. 

\subsection{Description of the Finite State Machine}\label{sec:punctCodesFSM}

When punctured convolutional codes are used, the strict relation between an
uncoded information bit, $n$ encoded bits and the mapped symbol is no longer
valid. As can be seen from Fig.~\ref{fig:convEncodingPunctured} (top), the
third and seventh encoded symbol, for example, are punctured and do not
contribute to the mapping process. Additionally, there is no strict relation
between MSB, LSB and the output of the generator polynomials, anymore. The
transitions of the FSM, depicted in Fig.~\ref{fig:convEncodingPunctured}
(bottom), show three differences when compared to \emph{non-punctured} FSM
transistions, \cf\ Fig.~\ref{fig:convEncodingTranditional} (bottom). These are
described subsequently.

\begin{figure}[ht]\vspace*{-2ex}
 \begin{center}
  \begin{tikzpicture}[>=latex,x=6em,y=4ex,font=\footnotesize,inner sep=0.3em,
                      node distance=10mm and 4mm]
   \coordinate (u0) at (0,0);
   \coordinate (u1) at (1,0);
   \coordinate (u2) at (2,0);
   \coordinate (u3) at (3,0);
   \draw (u0) circle (1pt) node[above] {$u[\nu]$};
   \draw (u1) circle (1pt) node[above] {$u[\nu+1]$};
   \draw (u2) circle (1pt) node[above] {$u[\nu+2]$};
   \draw (u3) circle (1pt) node[above] {$u[\nu+3]$};
   \draw (-0.25,-1) circle (1pt) node[coordinate] (c0) {};
   \draw (+0.25,-1) circle (1pt) node[coordinate] (c1) {};
   \draw (+0.75,-1) circle (1pt) node[coordinate] (c2) {};
   \draw (+1.25,-1) circle (1pt) node[coordinate] (c3) {};
   \draw (+1.75,-1) circle (1pt) node[coordinate] (c4) {};
   \draw (+2.25,-1) circle (1pt) node[coordinate] (c5) {};
   \draw (+2.75,-1) circle (1pt) node[coordinate] (c6) {};
   \draw (+3.25,-1) circle (1pt) node[coordinate] (c7) {};
   \begin{scope}[shorten <=1pt,shorten >=1pt]
    \draw[->] (u0) -- node[midway,above,sloped] {$g_1$} (c0);
    \draw[->] (u0) -- node[midway,above,sloped] {$g_2$} (c1);
    \draw[->] (u1) -- node[midway,above,sloped] {$g_1$} (c2);
    \draw[->] (u1) -- node[midway,above,sloped] {$g_2$} (c3);
    \draw[->] (u2) -- node[midway,above,sloped] {$g_1$} (c4);
    \draw[->] (u2) -- node[midway,above,sloped] {$g_2$} (c5);
    \draw[->] (u3) -- node[midway,above,sloped] {$g_1$} (c6);
    \draw[->] (u3) -- node[midway,above,sloped] {$g_2$} (c7);
    \draw (-0.25,-2) node[draw] (s0) {MSB};
    \draw (+0.25,-2) node[draw] (s1) {LSB};
    \draw (+0.75,-2) node       (s2) {\mbox{ }};
    \draw (+1.25,-2) node[draw] (s3) {MSB};
    \draw (+1.75,-2) node[draw] (s4) {LSB};
    \draw (+2.25,-2) node[draw] (s5) {MSB};
    \draw (+2.75,-2) node       (s6) {\mbox{ }};
    \draw (+3.25,-2) node[draw] (s7) {LSB};
   \end{scope}
   \node at (c2) {\Large\textbf{$\times$}};
   \node at (c6) {\Large\textbf{$\times$}};
   \begin{scope}[shorten <=1pt,shorten >=1pt]
    \foreach \x in {0,1,3,4,5,7} {
     \draw[->] (c\x) -- (s\x);
    }
   \end{scope}
   \draw ($(s0.south west)-(1pt,1pt)$) -| ($(s1.north east)+(1pt,1pt)$) -| ($(s0.south west)-(1pt,1pt)$);
   \draw ($(s3.south west)-(1pt,1pt)$) -| ($(s4.north east)+(1pt,1pt)$) -| ($(s3.south west)-(1pt,1pt)$);
   \draw ($(s5.south west)-(1pt,1pt)$) -| ($(s7.north east)+(1pt,1pt)$) -| ($(s5.south west)-(1pt,1pt)$);
   \begin{scope}[decoration={brace,amplitude=.5em},decorate]
    \draw[decorate] ($(s1.south east)-(0,2pt)$) -- node[yshift=-1ex,midway,below] {$m[\nu]$} ($(s0.south west)-(0,2pt)$);
    \draw[decorate] ($(s4.south east)-(0,2pt)$) -- node[yshift=-1ex,midway,below] {$m[\nu+1]$} ($(s3.south west)-(0,2pt)$);
    \draw[decorate] ($(s7.south east)-(0,2pt)$) -- node[yshift=-1ex,midway,below] {$m[\nu+2]$} ($(s5.south west)-(0,2pt)$);
   \end{scope}
  \end{tikzpicture}
  \vspace{2ex}
  \begin{tikzpicture}
   \tikzset{row 1 column 8/.style={nodes={STATE}}}
   \tikzset{row 1 column 7/.style={nodes={STATE}}}
   \tikzset{row 1 column 6/.style={nodes={INPUT}}}
   \tikzset{row 1 column 5/.style={nodes={EXTEND}}}
   \tikzset{row 2 column 7/.style={nodes={STATE}}}
   \tikzset{row 2 column 6/.style={nodes={STATE}}}
   \tikzset{row 2 column 5/.style={nodes={INPUT}}}
   \tikzset{row 2 column 4/.style={nodes={EXTEND}}}
   \tikzset{row 3 column 6/.style={nodes={STATE}}}
   \tikzset{row 3 column 5/.style={nodes={STATE}}}
   \tikzset{row 3 column 4/.style={nodes={INPUT}}}
   \tikzset{row 3 column 3/.style={nodes={EXTEND}}}
   \tikzset{row 4 column 4/.style={nodes={STATE}}}
   \tikzset{row 4 column 3/.style={nodes={STATE}}}
   \tikzset{row 4 column 2/.style={nodes={INPUT}}}
   \tikzset{row 4 column 1/.style={nodes={EXTEND}}}
   \matrix (FIFO) [matrix of nodes,
                   nodes in empty cells,
                   nodes={draw,
                          ultra thin,
                          anchor=south,
                          rectangle,
                          text width=2em,
                          minimum height=7ex,
                         },
                   ] {
     &&&&&&&\\
     &&&&&&&\\
     &&&&&&&\\
     &&&&&&&\\
    };
    \draw[XSB] ($(FIFO-1-6.west)+(0,1ex)$) -- node[above,font=\footnotesize,midway] {$g_1$} ($(FIFO-1-8.east)+(0,1ex)$);
    \draw[XSB] ($(FIFO-1-6.west)-(0,1ex)$) -- node[below,font=\footnotesize,midway] {$g_2$} ($(FIFO-1-8.east)-(0,1ex)$);
    \draw[XSB] ($(FIFO-2-5.west)+(0,1ex)$) -- node[above,font=\footnotesize,midway] {$g_2$} ($(FIFO-2-7.east)+(0,1ex)$);
    \draw[XSB] ($(FIFO-2-4.west)-(0,1ex)$) -- node[below,font=\footnotesize,midway] {$g_1$} ($(FIFO-2-6.east)-(0,1ex)$);
    \draw[XSB] ($(FIFO-3-4.west)+(0,1ex)$) -- node[above,font=\footnotesize,midway] {$g_2$} ($(FIFO-3-6.east)+(0,1ex)$);
    \draw[XSB] ($(FIFO-3-3.west)-(0,1ex)$) -- node[below,font=\footnotesize,midway] {$g_2$} ($(FIFO-3-5.east)-(0,1ex)$);
    \draw[XSB] ($(FIFO-4-2.west)+(0,1ex)$) -- node[above,font=\footnotesize,midway] {$g_1$} ($(FIFO-4-4.east)+(0,1ex)$);
    \draw[XSB] ($(FIFO-4-2.west)-(0,1ex)$) -- node[below,font=\footnotesize,midway] {$g_2$} ($(FIFO-4-4.east)-(0,1ex)$);
    \node[left] at (FIFO-1-1.west) {$\Gamma_0$};
    \node[left] at (FIFO-2-1.west) {$\Gamma_1$};
    \node[left] at (FIFO-3-1.west) {$\Gamma_2$};
    \node[left] at (FIFO-4-1.west) {$\Gamma_0$};
    \node[font=\footnotesize,right=1pt,draw,inner sep=1pt] at ($(FIFO-1-8.east)+(0,1.5ex)$)  {MSB};
    \node[font=\footnotesize,right=1pt,draw,inner sep=1pt] at ($(FIFO-1-8.east)+(0,-1.5ex)$) {LSB};
    \node[font=\footnotesize,right=1pt,draw,inner sep=1pt] at ($(FIFO-2-7.east)+(0,1.5ex)$)  {MSB};
    \node[font=\footnotesize,right=1pt,draw,inner sep=1pt] at ($(FIFO-2-7.east)+(0,-1.5ex)$) {LSB};
    \node[font=\footnotesize,right=1pt,draw,inner sep=1pt] at ($(FIFO-3-6.east)+(0,1.5ex)$)  {MSB};
    \node[font=\footnotesize,right=1pt,draw,inner sep=1pt] at ($(FIFO-3-6.east)+(0,-1.5ex)$) {LSB};
    \node[font=\footnotesize,right=1pt,draw,inner sep=1pt] at ($(FIFO-4-4.east)+(0,1.5ex)$)  {MSB};
    \node[font=\footnotesize,right=1pt,draw,inner sep=1pt] at ($(FIFO-4-4.east)+(0,-1.5ex)$) {LSB};
    \node[above,font=\footnotesize] at (FIFO-1-1.north) {$\nu+5$};
    \node[above,font=\footnotesize] at (FIFO-1-2.north) {$\nu+4$};
    \node[above,font=\footnotesize] at (FIFO-1-3.north) {$\nu+3$};
    \node[above,font=\footnotesize] at (FIFO-1-4.north) {$\nu+2$};
    \node[above,font=\footnotesize] at (FIFO-1-5.north) {$\nu+1$};
    \node[above,font=\footnotesize] at (FIFO-1-6.north) {$\nu$};
    \node[above,font=\footnotesize] at (FIFO-1-7.north) {$\nu-1$};
    \node[above,font=\footnotesize] at (FIFO-1-8.north) {$\nu-2$};
  \end{tikzpicture}\vspace*{-4ex}
 \end{center}
 \caption{Top: Encoding process for a rate-$\nicefrac23$ punctured convolutional
          code and natural labeling. Overall transmission rate
          $R=\nicefrac43$. Bottom: State transitions of the
          transmitter FSM and the relations between generator polynomials and
          FSM-state/input.}
 \label{fig:convEncodingPunctured}
 \vspace*{-2ex}
\end{figure}

\subsubsection{Generator Offsets $\Gamma_i$}

It becomes clear that the strict relation between the MSB, LSB and the
generator polynomial $g_1$ and $g_2$, respectively, no longer hold. The second
symbol, \ie, $m[\nu+1]$, for example, contains information about $u[\nu+1]$ and
$u[\nu+2]$.  In addition, the MSB is now generated by $g_2$ instead of $g_1$ as
was the case in the \emph{non-punctured} approach. Accordingly, the LSB, which
was generated by $g_2$ in the \emph{non-punctured} case, is now generated by
$g_1$.  It is also clear that the third symbol is generated by $u[\nu+2]$ and
$u[\nu+3]$ using the generator polynomial $g_2$ twice. 

To handle these relations we introduce a set of so called generator offsets
$\Gamma_i$ which describe, depending on the puncturing scheme, the relations
between generator polynomials, input value, FSM state, and mapping to MSB or
LSB, respectively. For example, $\Gamma_1$ indicates that the MSB output symbol
is generated with the input value (light gray), the FSM state, and the
generator polynomial $g_2$. On the other hand, to calculate the LSB we have to
use the extension (hatched block, \tikz \draw[EXTEND,baseline] rectangle
(1.5ex,1.5ex);\,) as input value so that $g_1$ is one step ahead of the LSB.
Therefore, the input value for the MSB is now part of the FSM state for the
LSB.
Note that, in the case of $M=4$, $\Gamma_0$ is used whenever no effect occurs
for puncturing, \ie, an even number of puncturings occured up to time instant
$\nu$, and the generator polynomials are synchronized with LSB and MSB.
$\Gamma_1$ is used, whenever an odd number of puncturings has happened and
$\Gamma_2$ is used when an additional puncturing synchronize the generator
polynomials with MSB and LSB again. The number of generator offsets needed to
describe all steps depends on the size of the modulation alphabet whereas the
generator offsets depend also on the puncturing scheme.

\subsubsection{State Extension}

The FSM transitions in Fig.~\ref{fig:convEncodingPunctured} (bottom) show that
the symbol $m[\nu+1]$ contains information on the uncoded information bits
$u[\nu+1]$, for which the MSB output was punctured, and the consecutive
information bit $u[\nu+2]$. We see that, when generating the output, the
calculation of the LSB is one step ahead to the MSB and considers one extra
information bit. This results in a trellis that has to be expanded (\ie,
splitted) by a factor of two. This can be easily seen from the figure as the
generator polynomials in the second step (generator offsets $\Gamma_1$) now
cover four blocks instead of just three.
However, for a $4$-ary transmission, when puncturing happens a second time, MSB
and LSB are resynchronized with the generator polynomials and a so called merge
happens in the trellis diagram.

\begin{figure}[ht]\vspace*{-2ex}
 \begin{center}
  \begin{tikzpicture}[x=25mm,y=4mm]
   \draw[fill] ( 1,7) circle ( 1pt) -- ( 2,7) circle ( 1pt);
   \draw[fill] ( 1,6) circle ( 1pt) -- ( 2,7) circle ( 1pt);
   \draw[fill] ( 1,5) circle ( 1pt) -- ( 2,6) circle ( 1pt);
   \draw[fill] ( 1,4) circle ( 1pt) -- ( 2,6) circle ( 1pt);
   \draw[fill] ( 1,7) circle ( 1pt) -- ( 2,5) circle ( 1pt);
   \draw[fill] ( 1,6) circle ( 1pt) -- ( 2,5) circle ( 1pt);
   \draw[fill] ( 1,5) circle ( 1pt) -- ( 2,4) circle ( 1pt);
   \draw[fill] ( 1,4) circle ( 1pt) -- ( 2,4) circle ( 1pt);
   \draw[fill] ( 2,7) circle ( 1pt) -- ( 3,7) circle ( 1pt);
   \draw[fill] ( 2,6) circle ( 1pt) -- ( 3,7) circle ( 1pt);
   \draw[fill] ( 2,5) circle ( 1pt) -- ( 3,6) circle ( 1pt);
   \draw[fill] ( 2,4) circle ( 1pt) -- ( 3,6) circle ( 1pt);
   \draw[fill] ( 2,3) circle ( 1pt) -- ( 3,5) circle ( 1pt);
   \draw[fill] ( 2,2) circle ( 1pt) -- ( 3,5) circle ( 1pt);
   \draw[fill] ( 2,1) circle ( 1pt) -- ( 3,4) circle ( 1pt);
   \draw[fill] ( 2,0) circle ( 1pt) -- ( 3,4) circle ( 1pt);
   \draw[fill] ( 2,7) circle ( 1pt) -- ( 3,3) circle ( 1pt);
   \draw[fill] ( 2,6) circle ( 1pt) -- ( 3,3) circle ( 1pt);
   \draw[fill] ( 2,5) circle ( 1pt) -- ( 3,2) circle ( 1pt);
   \draw[fill] ( 2,4) circle ( 1pt) -- ( 3,2) circle ( 1pt);
   \draw[fill] ( 2,3) circle ( 1pt) -- ( 3,1) circle ( 1pt);
   \draw[fill] ( 2,2) circle ( 1pt) -- ( 3,1) circle ( 1pt);
   \draw[fill] ( 2,1) circle ( 1pt) -- ( 3,0) circle ( 1pt);
   \draw[fill] ( 2,0) circle ( 1pt) -- ( 3,0) circle ( 1pt);
   \draw[fill] ( 3,7) circle ( 1pt) -- ( 4,7) circle ( 1pt);
   \draw[fill] ( 3,6) circle ( 1pt) -- ( 4,7) circle ( 1pt);
   \draw[fill] ( 3,5) circle ( 1pt) -- ( 4,7) circle ( 1pt);
   \draw[fill] ( 3,4) circle ( 1pt) -- ( 4,7) circle ( 1pt);
   \draw[fill] ( 3,3) circle ( 1pt) -- ( 4,6) circle ( 1pt);
   \draw[fill] ( 3,2) circle ( 1pt) -- ( 4,6) circle ( 1pt);
   \draw[fill] ( 3,1) circle ( 1pt) -- ( 4,6) circle ( 1pt);
   \draw[fill] ( 3,0) circle ( 1pt) -- ( 4,6) circle ( 1pt);
   \draw[fill] ( 3,7) circle ( 1pt) -- ( 4,5) circle ( 1pt);
   \draw[fill] ( 3,6) circle ( 1pt) -- ( 4,5) circle ( 1pt);
   \draw[fill] ( 3,5) circle ( 1pt) -- ( 4,5) circle ( 1pt);
   \draw[fill] ( 3,4) circle ( 1pt) -- ( 4,5) circle ( 1pt);
   \draw[fill] ( 3,3) circle ( 1pt) -- ( 4,4) circle ( 1pt);
   \draw[fill] ( 3,2) circle ( 1pt) -- ( 4,4) circle ( 1pt);
   \draw[fill] ( 3,1) circle ( 1pt) -- ( 4,4) circle ( 1pt);
   \draw[fill] ( 3,0) circle ( 1pt) -- ( 4,4) circle ( 1pt);
   \draw[shorten <=3pt,shorten >=3pt,|-|] (1,8) -- node[midway,circle,inner sep=2pt,fill=white,draw] {$\Gamma_0$} (2,8);
   \draw[shorten <=3pt,shorten >=3pt,|-|] (2,8) -- node[midway,circle,inner sep=2pt,fill=white,draw] {$\Gamma_1$} (3,8);
   \draw[shorten <=3pt,shorten >=3pt,|-|] (3,8) -- node[midway,circle,inner sep=2pt,fill=white,draw] {$\Gamma_2$} (4,8);
  \end{tikzpicture}\vspace*{-2ex}
 \end{center}
 \caption{Super-trellis representation for a punctured rate-$\nicefrac23$
          convolutional code. In the first to VA steps, two transitions arrive
          at each state, \ie, one bit can can be estimate, whereas the third
          step allows an estimation for two bits.}
 \label{fig:nonLinearTrellis}
 \vspace*{-2ex}
\end{figure}
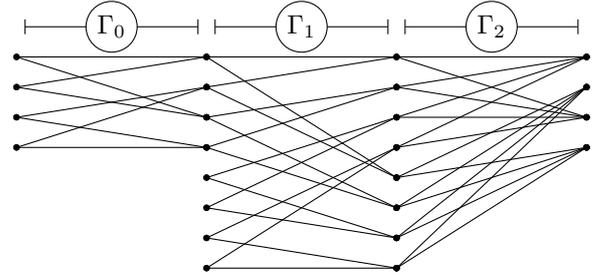

\begin{figure*}\vspace*{-3ex}
 \begin{center}
  \begin{tikzpicture}
   \tikzset{row 1 column 12/.style={nodes={STATE}}}
   \tikzset{row 1 column 11/.style={nodes={STATE}}}
   \tikzset{row 1 column 10/.style={nodes={STATE}}}
   \tikzset{row 1 column 9/.style={nodes={STATE}}}
   \tikzset{row 1 column 8/.style={nodes={INPUT}}}
   \tikzset{row 1 column 7/.style={nodes={EXTEND}}}
   \tikzset{row 2 column 10/.style={nodes={STATE}}}
   \tikzset{row 2 column 9/.style={nodes={STATE}}}
   \tikzset{row 2 column 8/.style={nodes={STATE}}}
   \tikzset{row 2 column 7/.style={nodes={STATE}}}
   \tikzset{row 2 column 6/.style={nodes={INPUT}}}
   \tikzset{row 2 column 5/.style={nodes={EXTEND}}}
   \tikzset{row 3 column 9/.style={nodes={STATE}}}
   \tikzset{row 3 column 8/.style={nodes={STATE}}}
   \tikzset{row 3 column 7/.style={nodes={STATE}}}
   \tikzset{row 3 column 6/.style={nodes={STATE}}}
   \tikzset{row 3 column 5/.style={nodes={INPUT}}}
   \tikzset{row 3 column 4/.style={nodes={EXTEND}}}
   \tikzset{row 4 column 8/.style={nodes={STATE}}}
   \tikzset{row 4 column 7/.style={nodes={STATE}}}
   \tikzset{row 4 column 6/.style={nodes={STATE}}}
   \tikzset{row 4 column 5/.style={nodes={STATE}}}
   \tikzset{row 4 column 4/.style={nodes={INPUT}}}
   \tikzset{row 4 column 3/.style={nodes={EXTEND}}}
   \tikzset{row 5 column 6/.style={nodes={STATE}}}
   \tikzset{row 5 column 5/.style={nodes={STATE}}}
   \tikzset{row 5 column 4/.style={nodes={STATE}}}
   \tikzset{row 5 column 3/.style={nodes={STATE}}}
   \tikzset{row 5 column 2/.style={nodes={INPUT}}}
   \tikzset{row 5 column 1/.style={nodes={EXTEND}}}
   \matrix (FIFO) [matrix of nodes,
                   nodes in empty cells,
                   nodes={draw,
                          ultra thin,
                          anchor=south,
                          rectangle,
                          text width=2em,
                          minimum height=7ex,
                         },
                   ] {
     &&&&&&&&&&&\\
     &&&&&&&&&&&\\
     &&&&&&&&&&&\\
     &&&&&&&&&&&\\
     &&&&&&&&&&&\\
    };
    \draw[XSB] ($(FIFO-1-8.west)+(0,2ex)$)            -- ($(FIFO-1-10.east)+(0,2ex)$);
    \draw[XSB] ($(FIFO-1-7.west)+(0,1ex)$)            -- ($(FIFO-1-9.east)+(0,1ex)$);
    \draw[XSB,opacity=.5] ($(FIFO-1-9.west)-(0,1ex)$) -- ($(FIFO-1-11.east)-(0,1ex)$);
    \draw[XSB,opacity=.5] ($(FIFO-1-8.west)-(0,2ex)$) -- ($(FIFO-1-10.east)-(0,2ex)$);
    \draw[XSB] ($(FIFO-2-6.west)+(0,2ex)$)            -- ($(FIFO-2-8.east)+(0,2ex)$);
    \draw[XSB] ($(FIFO-2-6.west)+(0,1ex)$)            -- ($(FIFO-2-8.east)+(0,1ex)$);
    \draw[XSB,opacity=.5] ($(FIFO-2-8.west)-(0,1ex)$) -- ($(FIFO-2-10.east)-(0,1ex)$);
    \draw[XSB,opacity=.5] ($(FIFO-2-7.west)-(0,2ex)$) -- ($(FIFO-2-9.east)-(0,2ex)$);
    \draw[XSB] ($(FIFO-3-5.west)+(0,2ex)$)            -- ($(FIFO-3-7.east)+(0,2ex)$);
    \draw[XSB] ($(FIFO-3-4.west)+(0,1ex)$)            -- ($(FIFO-3-6.east)+(0,1ex)$);
    \draw[XSB,opacity=.5] ($(FIFO-3-6.west)-(0,1ex)$) -- ($(FIFO-3-8.east)-(0,1ex)$);
    \draw[XSB,opacity=.5] ($(FIFO-3-6.west)-(0,2ex)$) -- ($(FIFO-3-8.east)-(0,2ex)$);
    \draw[XSB] ($(FIFO-4-4.west)+(0,2ex)$)            -- ($(FIFO-4-6.east)+(0,2ex)$);
    \draw[XSB] ($(FIFO-4-3.west)+(0,1ex)$)            -- ($(FIFO-4-5.east)+(0,1ex)$);
    \draw[XSB,opacity=.5] ($(FIFO-4-5.west)-(0,1ex)$) -- ($(FIFO-4-7.east)-(0,1ex)$);
    \draw[XSB,opacity=.5] ($(FIFO-4-4.west)-(0,2ex)$) -- ($(FIFO-4-6.east)-(0,2ex)$);
    \draw[XSB] ($(FIFO-5-2.west)+(0,2ex)$)            -- ($(FIFO-5-4.east)+(0,2ex)$);
    \draw[XSB] ($(FIFO-5-2.west)+(0,1ex)$)            -- ($(FIFO-5-4.east)+(0,1ex)$);
    \draw[XSB,opacity=.5] ($(FIFO-5-4.west)-(0,1ex)$) -- ($(FIFO-5-6.east)-(0,1ex)$);
    \draw[XSB,opacity=.5] ($(FIFO-5-3.west)-(0,2ex)$) -- ($(FIFO-5-5.east)-(0,2ex)$);
    \node[font=\footnotesize,right=1pt,ellipse,draw,inner sep=0pt]            at ($(FIFO-1-12.east)+(0,1.5ex)$) {$h[0]$};
    \node[font=\footnotesize,right=1pt,ellipse,draw,inner sep=0pt,opacity=.5] at ($(FIFO-1-12.east)+(0,-1.5ex)$) {$h[1]$};
    \node[font=\footnotesize,right=1pt,ellipse,draw,inner sep=0pt]            at ($(FIFO-2-10.east)+(0,1.5ex)$) {$h[0]$};
    \node[font=\footnotesize,right=1pt,ellipse,draw,inner sep=0pt,opacity=.5] at ($(FIFO-2-10.east)+(0,-1.5ex)$) {$h[1]$};
    \node[font=\footnotesize,right=1pt,ellipse,draw,inner sep=0pt]            at ($(FIFO-3-9.east)+(0,1.5ex)$) {$h[0]$};
    \node[font=\footnotesize,right=1pt,ellipse,draw,inner sep=0pt,opacity=.5] at ($(FIFO-3-9.east)+(0,-1.5ex)$) {$h[1]$};
    \node[font=\footnotesize,right=1pt,ellipse,draw,inner sep=0pt]            at ($(FIFO-4-8.east)+(0,1.5ex)$) {$h[0]$};
    \node[font=\footnotesize,right=1pt,ellipse,draw,inner sep=0pt,opacity=.5] at ($(FIFO-4-8.east)+(0,-1.5ex)$) {$h[1]$};
    \node[font=\footnotesize,right=1pt,ellipse,draw,inner sep=0pt]            at ($(FIFO-5-6.east)+(0,1.5ex)$) {$h[0]$};
    \node[font=\footnotesize,right=1pt,ellipse,draw,inner sep=0pt,opacity=.5] at ($(FIFO-5-6.east)+(0,-1.5ex)$) {$h[1]$};
    \node[left] at (FIFO-1-1.west) {\parbox{3em}{$\Gamma_2$\\$\hspace*{1em}\Gamma_1$}};
    \node[left] at (FIFO-2-1.west) {\parbox{3em}{$\Gamma_0$\\$\hspace*{1em}\Gamma_2$}};
    \node[left] at (FIFO-3-1.west) {\parbox{3em}{$\Gamma_1$\\$\hspace*{1em}\Gamma_0$}};
    \node[left] at (FIFO-4-1.west) {\parbox{3em}{$\Gamma_2$\\$\hspace*{1em}\Gamma_1$}};
    \node[left] at (FIFO-5-1.west) {\parbox{3em}{$\Gamma_0$\\$\hspace*{1em}\Gamma_2$}};
    \draw[densely dashed,line width=1pt] (FIFO-4-1.south west) -- (FIFO-4-12.south east);
    \draw[densely dashed,line width=1pt] (FIFO-1-1.south west) -- (FIFO-1-12.south east);
    \node[above,font=\footnotesize] at (FIFO-1-1.north) {$\nu+5$};
    \node[above,font=\footnotesize] at (FIFO-1-2.north) {$\nu+4$};
    \node[above,font=\footnotesize] at (FIFO-1-3.north) {$\nu+3$};
    \node[above,font=\footnotesize] at (FIFO-1-4.north) {$\nu+2$};
    \node[above,font=\footnotesize] at (FIFO-1-5.north) {$\nu+1$};
    \node[above,font=\footnotesize] at (FIFO-1-6.north) {$\nu$};
    \node[above,font=\footnotesize] at (FIFO-1-7.north) {$\nu-1$};
    \node[above,font=\footnotesize] at (FIFO-1-8.north) {$\nu-2$};
    \node[above,font=\footnotesize] at (FIFO-1-9.north) {$\nu-3$};
    \node[above,font=\footnotesize] at (FIFO-1-10.north){$\nu-4$};
    \node[above,font=\footnotesize] at (FIFO-1-11.north){$\nu-5$};
    \node[above,font=\footnotesize] at (FIFO-1-12.north){$\nu-6$};
    \path[fill=white,opacity=.6] (FIFO-1-1.north west) -| (FIFO-1-12.south east) -| (FIFO-1-1.north west);
    \path[fill=white,opacity=.6] (FIFO-5-1.north west) -| (FIFO-5-12.south east) -| (FIFO-5-1.north west);
  \end{tikzpicture}\vspace*{-2ex}
 \end{center}
 \caption{State transitions of the transmitter FSM with $R=\nicefrac43$ and the
          relations between generator polynomials, FSM-state/input and channel
          state for a memory-$1$ ISI-channel.}
 \label{fig:MatchedPuncturedConvISI}
 \vspace*{-2ex}
\end{figure*}
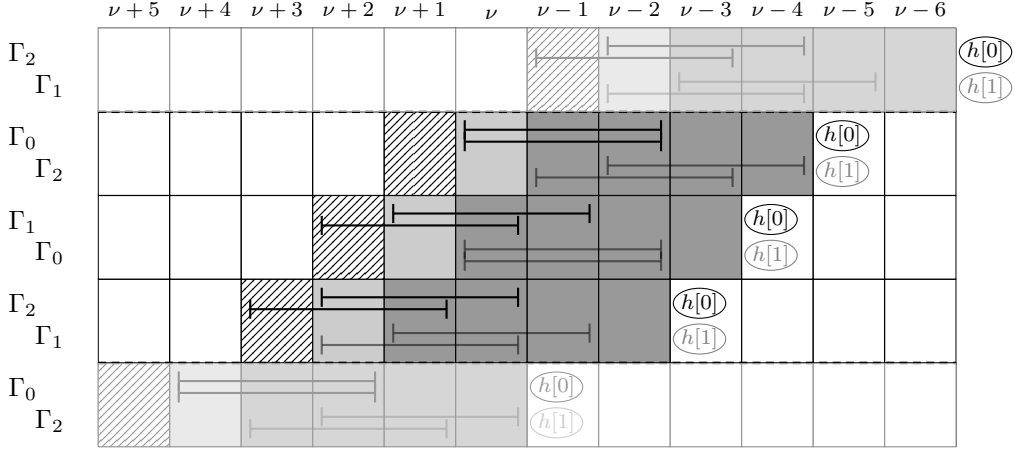

Therefore, a trellis based decoding algorithm, such as the VA, has to be
performed on a time-variant trellis diagram. As an example,
Fig.~\ref{fig:nonLinearTrellis} shows the resulting trellis diagram for a
punctured convolutional code with a constraint length $\nu=2$.

\subsubsection{Time Asynchronicity}

The punctured code has a rate of $R_\text{P}=\nicefrac23$. As we can not insert
Log-Likelihood Ratios (LLR) with $\mathcal{L}=0$ after equalization and before
decoding, because of the joint trellises of equalization and decoding, the VA
has to estimate four bits from just three received symbols to achieve the code
rate. As one can see from the trellis in Fig.~\ref{fig:nonLinearTrellis} the
first two steps $\Gamma_0$ have two transitions arriving at each state
resulting in an estimation of a single bit per state. However, the last step
$\Gamma_2$ has four transitions into each state so that the decision for a
survivor gives an estimate on two bits. The path register of the VA has to
consider the fact that now four bits have been estimated within three received
symbols.
The last step can be described as a state merging, whereas the split is
performed in the first step, \eg\ by copying the state metrics of the first
four states into the last four states.

\subsection{Viterbi Modifications for Punctured Codes}

As mentioned above the VA has to estimate four bits within three symbols from
the trellis transitions. To see the modifications of the VA we need to consider
the state extension.

Using Fig.~\ref{fig:convEncodingPunctured} (bottom), we define in each step
$\Gamma_0,\dots,\Gamma_2$ the input value to be the first value which is used
by either $g_1$ or $g_2$. As a result, in step $\Gamma_0$ the values at time
instants $k=\left\{ \nu;\;\nu+4 \right\}$ can be defined as input values.
When in step $\Gamma_1$ or $\Gamma_2$, the input value is at time instant
$k=\left\{ \nu+2;\; \nu+3\right\}$ (\eg, the hatched values, \tikz
\draw[EXTEND,baseline] rectangle (1.5ex,1.5ex);\,). Unfortunately we can not
estimate the input value $u[\nu+2]$ in $\Gamma_1$ because parts of the
information has not been received, yet, (\ie, missing information is located in
the MSB in $\Gamma_2$).
However, as all information on $u[\nu+1]$ (for which the output of $g_1$ was
punctured) has been received we can estimate it. To do so we have to evaluate
the most significant bit of the survivor state which is selected by the VA. In
$\Gamma_2$, as already mentioned, selecting the survivor path allows a decision
for two information bits (the most significant bit of the survivor \emph{and}
the input value) because four transitions end in each state, here.
For implementations, keep in mind that the VA is thus running asynchronously to
the desired information sequence at the output to achieve the overall rate
$R=\nicefrac43$.

\subsection{Matched Decoding}\label{sec:punctCodesME}

The above description now enables the extension for super-trellis decoding of
punctured convolutional coded transmission over ISI-channels.
To extend matched decoding for punctured codes, one must ensure that the right
generator offsets $\Gamma_i$ are involved at the right time.
Once the encoding and mapping is executed (here natural labeling and $4$-ASK
modulation are considered), the symbols traverse through the ISI-channel
independent from the encoding and puncturing process. For an ISI-channel with
two taps (channel memory $L=1$), for example, two output symbols of the mapper
are involved at each time instant. Thus, two out of the existing three
generator offsets $\Gamma_i$ and $\Gamma_{(i-1)\operatorname{mod}3}$ have to be
considered.

This can be seen from Fig.~\ref{fig:MatchedPuncturedConvISI}. For a memory-$1$
channel, in the third step (generator offset $\Gamma_2$), $m[\nu+2]$ is
received and the ISI-channel memory contains $m[\nu+1]$ which was generated in
the previous time instant using generator offset $\Gamma_1$.
The resulting overall generator offsets are depicted in
Fig.~\ref{fig:MatchedPuncturedConvISI}. This scheme can easily be extended to
arbitrary lengths of the ISI-channel.

\section{Numerical Results}\label{sec:numericalResults}

\begin{figure*}[!t]\vspace*{-3ex}
 \begin{subfigure}[t]{0.48\textwidth}
  \begin{center}
   \begin{tikzpicture}
    \begin{axis}[
                 width=6.5cm,
                 height=6cm,
                 xlabel={$10\log_{10}\left(\frac{E_\text{b}}{N_0}\right)$ in dB},
                 ylabel={BER},
                 enlargelimits=false,
                 ymode=log,
                 cycle list name=colors1Empty4Full,
                 grid=both,
                 xmin=2,xmax=15,
                 ymax=1,ymin=1e-4,
                 legend pos=outer north east,
                 every axis legend/.append style={font={\tiny},nodes={left}},
                 xtick={1,3,...,15},
                ]
     \addplot table[x index=0,y index=1]         {data_md-PAM_punctured-ST64-Senc4-Sfir16-Coded.data};
     \addplot table[x index=0,y index=2]         {data_md-PAM_punctured-ST64-Senc4-Sfir16-Coded.data};
     \addplot table[x index=0,y index=3]         {data_md-PAM_punctured-ST64-Senc4-Sfir16-Coded.data};
     \addplot table[x index=0,y index=4]         {data_md-PAM_punctured-ST64-Senc4-Sfir16-Coded.data};
     \addlegendentry{MD        (64 states)}
     \addlegendentry{DFSE-VA  (4+4 states)}
     \addlegendentry{DFSE-VA (16+4 states)}
     \addlegendentry{BCJR-VA (16+4 states)}
     \coordinate (note) at (rel axis cs:0.05,0.05);
    \end{axis}
    \node[draw,fill=white,anchor=south west,font={\footnotesize}] at (note) {$Z_\text{enc} = 2^4$, $Z_\text{cha}=4^2$};
   \end{tikzpicture}
  \end{center}
 \end{subfigure}
 \begin{subfigure}[t]{0.48\textwidth}
  \begin{center}
   \begin{tikzpicture}
    \begin{axis}[
                 width=6.5cm,
                 height=6cm,
                 xlabel={$10\log_{10}\left(\frac{E_\text{b}}{N_0}\right)$ in dB},
                 ylabel={BER},
                 enlargelimits=false,
                 ymode=log,
                 cycle list name=colors1Empty4Full,
                 grid=both,
                 xmin=2,xmax=15,
                 ymax=1,ymin=1e-4,
                 legend pos=outer north east,
                 every axis legend/.append style={font={\tiny},nodes={left}},
                 xtick={1,3,...,15},
                ]
     \addplot table[x index=0,y index=1]         {data_md-PAM_punctured-ST256-Senc4-Sfir64-Coded.data};
     \addplot table[x index=0,y index=2]         {data_md-PAM_punctured-ST256-Senc4-Sfir64-Coded.data};
     \addplot table[x index=0,y index=3]         {data_md-PAM_punctured-ST256-Senc4-Sfir64-Coded.data};
     \addplot table[x index=0,y index=4]         {data_md-PAM_punctured-ST256-Senc4-Sfir64-Coded.data};
     \addplot table[x index=0,y index=5]         {data_md-PAM_punctured-ST256-Senc4-Sfir64-Coded.data};
     \addlegendentry{MD       (128 states)}
     \addlegendentry{DFSE-VA  (4+4 states)}
     \addlegendentry{DFSE-VA (16+4 states)}
     \addlegendentry{DFSE-VA (64+4 states)}
     \addlegendentry{BCJR-VA (64+4 states)}
     \coordinate (note) at (rel axis cs:0.05,0.05);
    \end{axis}
    \node[draw,fill=white,anchor=south west,font={\footnotesize}] at (note) {$Z_\text{enc} = 2^2$, $Z_\text{cha}=4^3$};
   \end{tikzpicture}
  \end{center}
 \end{subfigure}\vspace*{-1ex}
 \caption{Bit error performance for convolutionally encoded $4$-ASK
          transmission (gray labeling) with overall rate $R =
          \nicefrac43$ over the ISI-AWGN-channel.  Generator polynomials
          $\left[ 5_\text{oct};\; 7_\text{oct}\right]$ ($Z_\text{cha}=4$).
           Left: Memory-$2$ ISI-channel ($L=2$, \ie, $Z_\text{cha}=16$).
          Right: Memory-$3$ ISI-channel ($L=3$, \ie, $Z_\text{cha}=64$).
  }
 \label{fig:BERSfir1664}
 \vspace*{-2ex}
\end{figure*}
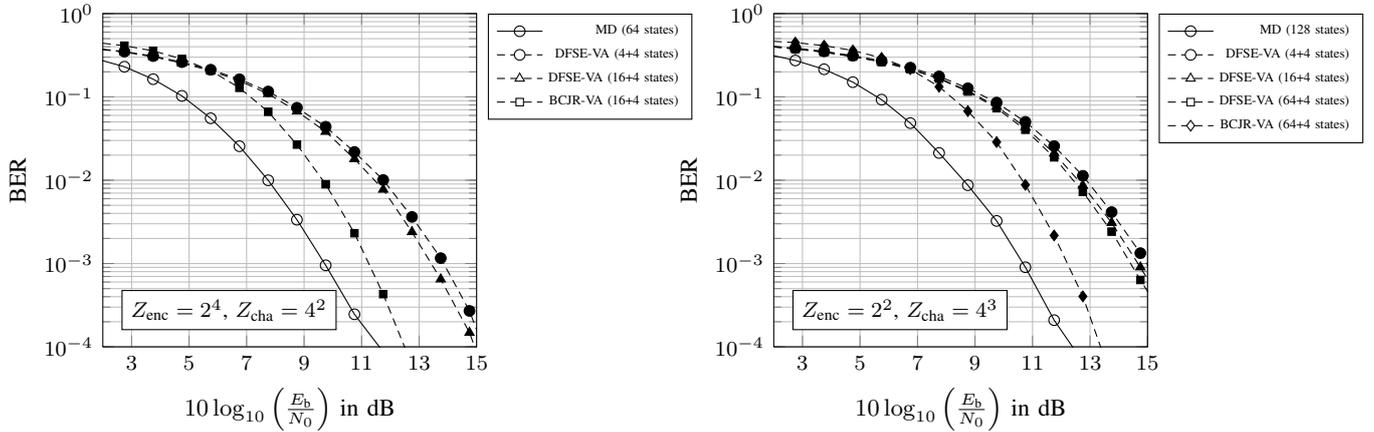

The effectiveness of the approach of punctured MD is now verified by means of
numerical simulations. We restrict ourselves to rate-$\nicefrac12$ encoding
schemes and a $4$-ary modulation alphabet. As convolutional encoder we apply
the generator polynomials $\left[5_\text{oct};\;7_\text{oct} \right]$ which, in
combination with Gray labeling, results in a trellis coded modulation scheme
(TCM) for $M$-ary ASK. The puncturing scheme is defined as $P_1 =
\left\{\;1;\;0\;\right\}$ and $P_2=\left\{\;1;\;1\;\right\}$ as shown in
Figure~\ref{fig:sysmodel} and~\ref{fig:convEncodingPunctured}. Thus, the overall
transmission rate is $R=\nicefrac43$. Please note, that this combination of
encoder, puncturing scheme, and mapper may not be the optimum choice. However,
it allows a comprehensible description of our approach.

For simplicity an exemplary minimum phase ISI-channel is generated by 
\begin{align}\label{eq:channelISI}
 h[k]     & = \frac{1}{\alpha}\;\cdot\;\frac{L-k+1}{L+1}; \qquad 0\leq k\leq L\\
 \alpha^2 & = \sum\limits_{k=0}^L\left( \frac{L-k+1}{L+1} \right)^2
\end{align}
and normalized to unit energy. Please note that due to the normalization the
equivalent energy per bit $E_\text{b}$ is identical at transmitter output and
receiver input. The applied ISI-channel is described by (\ref{eq:channelISI})
using $L\in\left\{ 2;\;3;\;4\right\}$ with $Z_\text{cha} =
\left\{16;\;64;\;256\right\}$ states, respectively.

The simulation results in Fig.~\ref{fig:BERSfir1664} shown the bit error rate
(BER) over an ISI-AWGN-channel with ratio of energy per information bit and
one-sided power spectral density $\frac{E_\text{b}}{N_0}$.

Our MD approach of the VA operating on the time-variant equivalent non-linear
trellis description is compared to separate equalization and decoding employing
DFSE/BCJR~\cite{BCJR74} for equalization and full-state VA for decoding. Please
note that by dispensing the interleaver between channel encoding and modulation
for the separated approaches, block errors that are caused by the equalization
process reduce the ability to decode due to correlated errors.
Obviously, the soft-decision separated approach results in improved bit error
rates when compared to to hard-decision separated approach. However, the
separated equalization and decoding approach is significantly outperformed by
MD as decoding is carried out in the super-trellis. Note that the results are
well-known for super-trellis decoding but are achieved with fewer states due to
the equivalent non-linear trellis description.

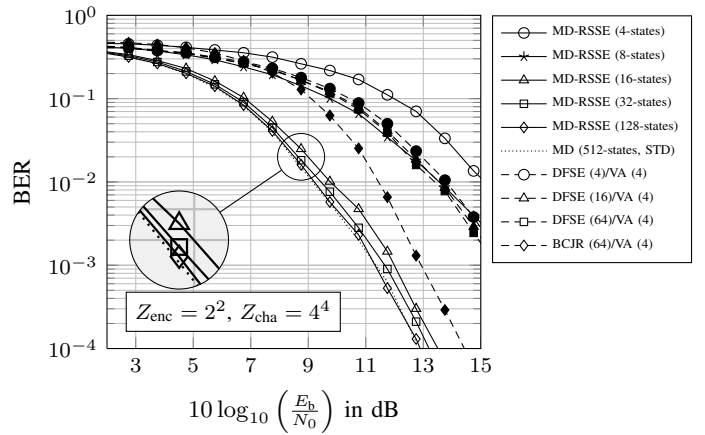
\begin{figure}[ht]\vspace*{-2ex}
 \begin{center}
  \begin{tikzpicture}[
                      spy using outlines={circle,magnification=2.1,connect spies},
                     ]
   \begin{axis}[
                width=6.5cm,
                height=6cm,
                xlabel={$10\log_{10}\left(\frac{E_\text{b}}{N_0}\right)$ in dB},
                ylabel={BER},
                enlargelimits=false,
                ymode=log,
                cycle list name=colors6Empty4Full,
                grid=both,
                xmin=2,xmax=15,
                ymax=1,ymin=1e-4,
                legend pos=outer north east,
                every axis legend/.append style={font={\tiny},nodes={right}},
                xtick={1,3,...,15},
               ]
    \addplot table[x index=0,y index=1]         {data_md-PAM_punctured-RSSE-ST1024-Senc4-Sfir256-Coded.data};
    \addplot table[x index=0,y index=2]         {data_md-PAM_punctured-RSSE-ST1024-Senc4-Sfir256-Coded.data};
    \addplot table[x index=0,y index=3]         {data_md-PAM_punctured-RSSE-ST1024-Senc4-Sfir256-Coded.data};
    \addplot table[x index=0,y index=4]         {data_md-PAM_punctured-RSSE-ST1024-Senc4-Sfir256-Coded.data};
    \addplot table[x index=0,y index=6]         {data_md-PAM_punctured-RSSE-ST1024-Senc4-Sfir256-Coded.data};
    \addplot[black,densely dotted] table[x index=0,y index=9]         {data_md-PAM_punctured-RSSE-ST1024-Senc4-Sfir256-Coded.data};
    \addplot table[x index=0,y index=10]        {data_md-PAM_punctured-RSSE-ST1024-Senc4-Sfir256-Coded.data};
    \addplot table[x index=0,y index=11]        {data_md-PAM_punctured-RSSE-ST1024-Senc4-Sfir256-Coded.data};
    \addplot table[x index=0,y index=12]        {data_md-PAM_punctured-RSSE-ST1024-Senc4-Sfir256-Coded.data};
    \addplot table[x index=0,y index=13]        {data_md-PAM_punctured-RSSE-ST1024-Senc4-Sfir256-Coded.data};
    \addlegendentry{MD-RSSE   (4-states)}
    \addlegendentry{MD-RSSE   (8-states)}
    \addlegendentry{MD-RSSE  (16-states)}
    \addlegendentry{MD-RSSE  (32-states)}
    \addlegendentry{MD-RSSE (128-states)}
    \addlegendentry{MD      (512-states, STD)}
    \addlegendentry{DFSE (4)/VA (4)}
    \addlegendentry{DFSE (16)/VA (4)}
    \addlegendentry{DFSE (64)/VA (4)}
    \addlegendentry{BCJR (64)/VA (4)}
    \coordinate (note) at (rel axis cs:0.05,0.05);
    \coordinate (spypoint) at (axis cs:8.75,2e-02);
   \end{axis}
   \node[draw,fill=white,anchor=south west,font={\footnotesize}] at (note) (nodeText) {$Z_\text{enc} = 2^2$, $Z_\text{cha}=4^4$};
   \spy[width=1.3cm,height=1.3cm] on (spypoint) in node [thin,fill=white!94!black,anchor=south,xshift=20pt,yshift=1pt] at (nodeText.north west);
  \end{tikzpicture}\vspace*{-2ex}
 \end{center}
 \caption{Bit error performance for convolutionally encoded $4$-ASK
          transmission (gray labeling) with overall rate $R =
          \nicefrac43$  over memory-$4$ the ISI-AWGN-channel
          ($Z_\text{cha}=256$) with Generator polynomials $\left[
          5_\text{oct};\; 7_\text{oct}\right]$ ($L=4$, \ie, $Z_\text{cha}=256$).
  }
 \label{fig:BERSfir256}
 \vspace*{-2ex}
\end{figure}

We also conducted simulations for larger trellises as shown in
Fig.~\ref{fig:BERSfir256}. There, the convolutional code can be described with
a four-state FSM and the ISI-channel has five taps resulting in $256$ states,
for a $4$-ary ASK-transmission. In the case of a \emph{non-punctured} code the
straigt-forward super-trellis has $1024$ states. The matched decoding approach
reduces the super-trellis to only $64$ states using a non-linear trellis
description~\cite{2012arXiv1207.4680S}.
However, for a punctured convolutional code, the time-variant super-trellis
would have $1024$ states, and $2048$ states when splitted. The matched decoding
approach allows a state reduction to $256$ and $512$ states, respectively.
Apparently, already for moderate encoder size and short ISI-channels,
super-trellis decoding becomes intractable, when the code is punctured due to
the state expansion.

Therefore, we also implemented an algorithm to perform the VA on a reduced set
of states. Due to lack of space we can not describe the modification to the
reduced-state sequence estimtion (RSSE) in full detail, here. The state
partitioning needed for RSSE uses the minimum-phase charateristic of the
impulse response of the ISI-channel and can be related to decision feed-back
sequence estimation, a partial solution of RSSE. A detailed description on the
application of RSSE of \emph{non-punctured} convolutional codes can be found
in~\cite{2012arXiv1207.4680S}.

The simulation results shown in Fig.~\ref{fig:BERSfir256} indicate, that our
MD-RSSE approach for punctured convolutional codes enables efficient
super-trellis decoding and also allows a trade-off between complexity and
performance, \ie, noise-robustness can be achieved. Obviously, the proposed
decoding scheme significantly supersedes the separated equalization and
decoding approaches already for only $16$ states.

\section{Conclusion}\label{sec:conclusion}

In this paper we have shown that it is possible to perform trellis decoding of
punctured convolutional encoded transmissions. We extended the well-known
decoding approach for the use over ISI-channels and described the differences
of matched decoding between non-punctured and punctured codes. By using RSSE
with DFSE-like partitioning we obtain an efficient method for a trade-off
between complexity and performance.

\IEEEtriggeratref{1}
\IEEEtriggercmd{\enlargethispage{1in}}
\bibliographystyle{IEEEtran}
\bibliography{IEEEabrv,main.bib}
\end{document}